\documentclass [floatfix,twocolumn,showpacs,preprintnumbers,amsmath,amssymb]{revtex4}
\usepackage{epsf,epsfig}
\usepackage[hypertex,linkcolor=red]{hyperref}

\newcommand{\Prob}{{\mathrm{Pr}}}

\newcommand{\E}{{\mathrm{e}}}
\newcommand{\I}{{\mathrm{i}}}

\newcommand{\ket}[1]{|#1\rangle}

\newcommand{\smfrac}[2]{\mbox{$\frac{#1}{#2}$}}
\newtheorem{procedure}{Procedure}

\begin{document}
\title{Optimal quantum circuits for general phase estimation}
\author{Wim van Dam$^{1}$, G. Mauro D'Ariano$^{2}$, Artur Ekert$^{3}$,
Chiara Macchiavello$^{2}$, and Michele
Mosca$^{4}$\footnote{Supported by DTO-ARO, NSERC, CFI, ORDCF, CIAR,
CRC, ORF, and Ontario-MTI}}
\address{$^3$Centre for Quantum Computation, DAMTP, University of Cambridge, United Kingdom}
\address{$^2$ Dipartimento di Fisica
``A.\,Volta'' and INFM - Unit\`a di Pavia, Via Bassi 6, I--27100
Pavia, Italy}
\address{$^1$Departments of Computer Science and Physics,
University of California, Santa Barbara, Santa Barbara, CA
93106-5110 United States of America}
\address{$^4$Institute for Quantum Computing, University of Waterloo
and St. Jerome's University, and Perimeter Institute for Theoretical
Physics, ON, Canada}
\begin{abstract}
We address the problem of estimating the phase $\phi$ given $N$
copies of the phase rotation gate $u_{\phi}$. We consider, for the
first time, the optimization of the general case where the circuit
consists of an arbitrary input state, followed by any arrangement of
the $N$ phase rotations interspersed with arbitrary quantum
operations, and ending with a POVM.  Using the polynomial method, we
show that, in all cases where the measure of quality of the estimate
$\tilde{\phi}$ for $\phi$ depends only on the difference
$\tilde{\phi} - \phi$, the optimal scheme has a very simple fixed
form. This implies that an optimal general phase estimation
procedure can be found by just optimizing the amplitudes of the
initial state.
\end{abstract}
\pacs{03.67.-a} \maketitle

The possibility of encoding information into the relative phase of
quantum systems is often exploited in several quantum information
processing tasks and several kinds of applications. For example, it
was shown that in most existing quantum algorithms the information
to be retrieved after the computation is contained in relative
phases \cite{CEMM}. Moreover, information is encoded into phase
properties in some quantum cryptographic protocols \cite{BB84}, and
in some precision measurements, such as the schemes on which atomic
clocks are based \cite{atclocks}. Therefore, the issue of estimating
the phase in the most efficient way is of great interest.

We phrase the phase estimation problem as follows. Let $u_{\phi}$ be
a single qubit gate that, in a prescribed `computational' basis
$\{\ket{0},\ket{1}\}$ maps the state $\ket{0}$ to $\ket{0}$ and
$\ket{1}$ to $\E^{\I\phi}\ket{1}$. We assume we have no prior
knowledge about $\phi$. The objective is to estimate $\phi$ using
some procedure that will output some guess $\tilde{\phi}$. We
characterize the quality of a estimate by a ``cost function''
$C(\phi, \tilde{\phi})$, which specifies the penalty associated with
guessing $\tilde{\phi}$ when the actual phase is $\phi$. We are
given $N$ identical single qubit quantum gates $u_\phi$, and the
goal is to use these gates along with any other operations in order
to produce an estimate of $\phi$. The optimal procedure is the one
that has the minimum expected cost.

Most of the previous work on phase estimation assumes some fixed
state encoding the phases, and the only thing to be optimized is the
final POVM measurement \cite{helstrom, holevo77, dbe}. More recent
work \cite{DDEMMb} fixes the way the phase gates are applied and
optimizes the choice of input state and final POVM.

The crucial point is that in this paper we are not restricted to
preparing some input state, then applying all of the phase
rotations, and then performing an optimal POVM. We consider, for the
first time, the case where one has full freedom over how to use the
phase rotation gates in an experiment designed to optimally estimate
the phase. Any realistic experiment of this type can be viewed as
computation, completely specified by a quantum circuit acting on
some finite number of qubits and involving, apart from the $N$
copies of the $u_\phi$ gates, some finite number of arbitrary
quantum gates of our choice. In fact, many quantum algorithms,
including Shor's quantum algorithm for factoring integers, can be
phrased in terms of such phase estimations~\cite{kitaev,CEMM}. This
has originally provided motivation for this work.

We assume the phase $\phi$ is chosen uniformly from $[0, 2 \pi)$
\footnote{
A uniform prior is also relevant in other scenarios, for
example, if we are working in an adversarial scenario where Alice
fixes her approximation scheme and then Bob (the adversary) picks
the phase in order maximize the expected cost of Alice's estimate.
Regardless of Bob's strategy, Alice can ``uniformize'' it by adding
a uniform (or arbitrarily close to uniform) random phase shift to
whatever phase shift gate is provided by Bob.  This means that any
adversary is no more powerful than an adversary that picks a phase
uniformly at random.}  and that a suitable quantum circuit
containing $N$ copies of the $u_\phi$ gates outputs some value $y$
with probability $\Prob(y|\phi)$. From $y$ we infer, following a
prescribed rule, the estimate $\tilde{\phi}_y$. The quality of the
whole procedure is quantified by the expected cost $\bar{C}$, given
by
\begin{equation}
\bar{C} =  \frac{1}{2\pi} \sum_y {{\int_{\phi=0}^{2\pi} {d\phi
\Prob(y|\phi)\cdot C(\phi,\tilde{\phi}_y)  }}}.
\end{equation}

The next part of this paper describes a very simple procedure for
estimating $\phi$, that only requires one to optimize the choice of
initial state to an otherwise fixed procedure. The rest of the paper
then reduces the very general case we have described above to this
very simple case.

For a given cost function, the quality of an estimation procedure
depends on both $\Prob(y|\phi)$ and the inference rule $y
\mapsto\tilde{\phi}_y$. The optimal protocol gives the minimum
possible average $\bar{C}$.  We restrict attention to cost functions
$C$ that depend only on $\phi - \tilde{\phi}_y$, and therefore adopt
the notation $C(\phi,\tilde{\phi}_y) = C(\phi-\tilde{\phi}_y)$. We
will make only the following very weak assumption on the cost
function (which corresponds to a more general class of cost
functions  than the ``Holevo'' class that is typically considered
\cite{holevo})
\begin{equation} \label{e:general-cost}
\int_{\phi=0}^{2\pi} {d\phi |C(\phi)|  } < \infty.
\end{equation}
We will deal with specific cost functions later on.

\begin{figure}
\begin{center}
\epsfig{file=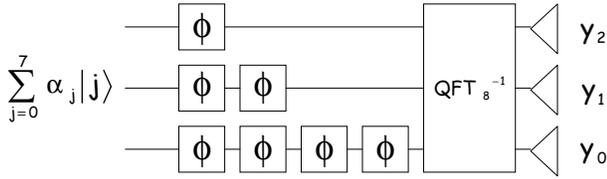,angle=-90,width=8cm} \vspace{3mm}
\end{center}
\caption{A simple approach for estimating $\phi$, in the case that
$N=7$. Optimizing over the input amplitudes $\alpha_j$, produces an
optimal estimate of $\phi$. }
\label{optimal_fig}
\end{figure}

Let us start by describing a simple and natural approach for
estimating the phase $\phi$, illustrated in Figure
\ref{optimal_fig}.

\begin{procedure}  \label{optimal_procedure}

\begin{itemize}

\item Prepare $m$ qubits in state
$\ket{x}=\sum_{j=0}^{N}\alpha_j\ket{j}$, with $N\leq 2^m-1$.
The exact values of
$\alpha_j$ depend on the cost function to be maximized.
\item Apply the $u_\phi$ gates to effect $U_\phi$
\begin{equation}
U_\phi\sum_{j=0}^{N}\alpha_j\ket{j} = \sum_{j=0}^{N}\alpha_j\E^{\I
j\phi}\ket{j}.
\end{equation}
\item Apply the inverse quantum Fourier transform to obtain
\begin{equation}
2^{-m/2}\sum_{y=0}^{2^m-1} \left(\sum_{j=0}^N\alpha_j\E^{\I
j(\phi-\smfrac{2\pi y}{2^m})} \right)\ket{y},
\end{equation}
measure $y$ and calculate the estimate $\tilde{\phi}_y=
\smfrac{2\pi y}{2^m}$.
\end{itemize}

\end{procedure}

\begin{figure}
\begin{center}
\epsfig{file=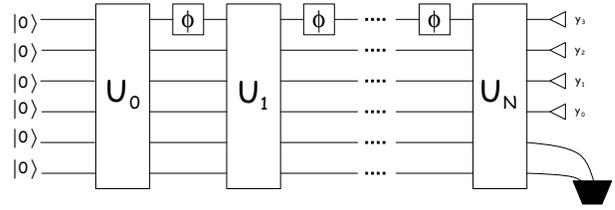,angle=-90,width=8cm} \vspace{3mm}
\caption{Most general approach for estimating the unknown phase
rotation. This subsumes Procedure \ref{optimal_procedure}, as well
as more complicated procedures with classical feedback.}
\label{general_phase_est_fig}
\end{center}
\end{figure}

The surprising claim is the following.  Given \emph{any} function
$C$ satisfying Eq. (\ref{e:general-cost}), the minimum of $\bar{C}$
obtained by optimizing the $\alpha_j$ in Procedure
\ref{optimal_procedure} is the infimum of all values obtainable by
\emph{any} realistic experiment (as we described above and
illustrate in Figure \ref{general_phase_est_fig}). It is important
to also note that apart from the preparation of the initial state,
the above procedure can be implemented using the $N$ black boxes
$u_\phi$ and a number of elementary gates polynomial in $\log(N)$.
Efficient preparation of states is discussed in \cite{km}. Exact
implementation of quantum Fourier transforms is discussed in
\cite{mz}, and arbitrarily good approximations are discussed in
\cite{kitaev, hh}.

The remainder of this paper will prove this claim by a sequence of
reductions.

 Let us start with a very general circuit (Figure
\ref{general_phase_est_fig}) which uses $m+d$ qubits, where $m$ and
$d$ can be arbitrarily large. The first $m$ qubits are measured
after the computation, yielding the output $0\leq y\leq 2^m-1$,
whereas the remaining $d$ qubits are discarded.

Since we are allowing arbitrarily many extra ``ancilla'' qubits, and
since any classical feedback scheme can in principle be implemented
by a unitary operation using a sufficiently large ancilla, the
family of schemes that can be implemented by a quantum circuit of
this form includes any scheme using finite dimensional state spaces.

For convenience, we let the output $y$ correspond to the phase
estimate $\tilde{\phi}_y = \smfrac{2\pi y}{M}$, for some $M$, $M\le
2^m$. For a fixed approximation scheme (using finite means) and cost
function  satisfying Eq. (\ref{e:general-cost}), and assuming a
uniform prior distribution of the $\phi$, this simplifying
assumption will give us a scheme with expected cost that is at most
$\bar{C} + \epsilon_M$, where $\epsilon_M \rightarrow 0$ as $M
\rightarrow \infty$, and $\bar{C}$ is the lowest expected cost for
any possible scheme. Thus the infimum of the $\bar{C}$ over all such
restricted schemes equals the infimum of the $\bar{C}$ over all
possible such schemes.

In fact we will also show later that as long as $M \geq N+1$, this
simplifying assumption does not cost us anything. That is, the
infimum of the expected costs of all the schemes using the inference
rule $y \mapsto \frac{2 \pi y}{M}$ for any $M \geq N+1$ is the
infimum of the possible expected costs using {\it any} inference
rule $y \mapsto \tilde{\phi}_y$.

Suppose we came up with a general circuit that performs an
estimation of $\phi$ according to some prescribed set of criteria.
Let us first show that such a circuit is equivalent, for our
purposes, to another one, which has much simpler structure.

\begin{figure}
\begin{center}
\epsfig{file=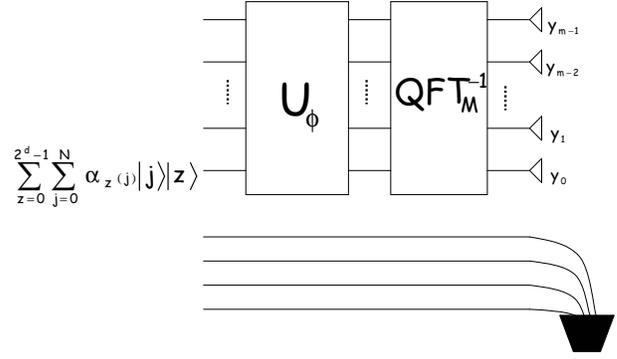,angle=-90,width=8cm} \vspace{3mm}
\caption{Without loss of generality, we can assume our phase
estimation procedure has the form illustrated in this figure, where
$2^m \geq M \geq N+1$. The top register contains $m$ qubits, and the
bottom $d$ qubits are ancilla qubits that may be entangled with the
first $m$ qubits, but are discarded.} \label{intermediate-fig}
\end{center}
\end{figure}

The state at the output of the circuit can be written as
\begin{equation}
\sum_{y=0}^{M}\sum_{z=0}^{2^d-1}\alpha(y,z,\phi)\ket{y}\ket{z},
\end{equation}
where each amplitude $\alpha(y,z,\phi)$ is a polynomial in
$\E^{\I\phi}$ of degree at most $N$
\begin{equation}
\alpha(y,z,\phi)=\sum_{j=0}^N \frac{\alpha_j (y,z)}{\sqrt{M}}
\E^{\I j\phi}.
\end{equation}
This fact follows just as in \cite{BBCMW} where the polynomial
method is applied to an oracle revealing one of many Boolean
variables. This paper is the first time the polynomial method is
applied to oracles with continuous variables.
Since we assume that the cost function is of the form
$C(\phi,\tilde{\phi}_y) = C(\phi-\tilde{\phi}_y)$, then we can also
assume, without loss of generality  that the optimal conditional
probability
\begin{equation}
\Prob(y|\phi) =\sum_z |\alpha(y,z,\phi)|^2
\end{equation}
depends only on the difference $\phi-\smfrac{2\pi y}{M}$  and
therefore equals
\begin{equation}
\Prob(0|\phi - 2 \pi y/M ) =\sum_z |\alpha(0,z,\phi - 2 \pi
y/M)|^2.
\end{equation}
To simplify the notation, we let $\alpha_j(z) = \alpha_j(0,z)$.
Therefore,  a circuit that produces amplitudes
\begin{equation}
|\alpha(y,z,\phi)|=\frac{1}{M}|\sum_{j=0}^N
\alpha_j (z) \E^{\I j(\phi-\smfrac{2\pi y}{M})}|
\end{equation}
also leads to the optimal estimation of $\phi$. Thus the following
simple estimation procedure, whose circuit is illustrated in Figure
\ref{intermediate-fig}, performs equally well:
\begin{itemize}
\item Prepare $m+d$ qubits in state
$\ket{x}=\sum_{z=0}^{2^d-1}\sum_{j=0}^{N}\alpha_j(z)\ket{j}\ket{z}$.
For this preparation to be possible $m$ has to be chosen such that
$N< 2^m$.
\item Apply the $u_\phi$ gates to effect $U_\phi$ on the first $m$ qubits
\begin{eqnarray}
U^N_\phi\sum_{z=0}^{2^d-1}\sum_{j=0}^{N}\alpha_j (z)\ket{j}\ket{z}
& = &\sum_{z=0}^{2^d-1}\sum_{j=0}^{N}\alpha_j (z)\E^{\I
j\phi}\ket{j}\ket{z}.
\end{eqnarray}
\item Apply the inverse quantum Fourier transform, $\ket{j}
\mapsto (1/{\sqrt M})\sum_{y=0}^{M-1}{\E^{-\I \smfrac{2\pi
y}{M}}\ket{y}}$, to obtain
\begin{equation}\label{eq:final_state}
\frac{1}{\sqrt M}\sum_{z=0}^{2^d-1}\sum_{y=0}^{M-1}
\left(\sum_{j=0}^N\alpha_j (z)\E^{\I j(\phi-\smfrac{2\pi y}{M})}
\right)\ket{y}\ket{z},
\end{equation}
and measure $y$.
\end{itemize}
The following two observations lead to further simplifications.

Let us first notice that in this procedure the role of the $d$
auxiliary qubits is restricted to the initial preparation of the
most general state of the first $m$ qubits (all subsequent
operations are restricted to these $m$ qubits). Usually, such a
state is described by a density operator which can always be
expressed as a mixture of pure states of the first $m$ qubits. The
average cost in this case is given by the average, over the mixture,
of individual expected costs pertaining to the pure states in the
mixture. Thus the expected cost for the mixture cannot be greater
than all of the individual expected costs for the contributing pure
states; hence either some of the contributing costs are greater or
they are all equal. In either case a judicial choice of a pure state
of the $m$ qubits leads to equally good or better phase estimation.
This argument implies that without loss of generality we can
restrict our circuit to only $m$ qubits (plus some ancilla bits used
to implement $U_\phi^N$ using $N$ copies of $u_\phi$) and run the
estimation on pure states.

Secondly, in the description above the quantum Fourier transform is
parameterized by $M$, where $N+1\le M\le 2^m$, but in fact any $M
\ge N+1$, in particular $M=2^m$, will work equally well. To see this
consider a cost function of the form $C(\phi-\tilde{\phi}_y)$. The
expected cost, for some $M$, is
\begin{eqnarray*}
\bar{C} & = &  \frac{1}{2\pi M}{{\int_{\phi=0}^{2\pi} d\phi
\sum_{y=0}^{M-1}{ \left|\sum_{j=0}^N\alpha_j \E^{\I
j(\phi-\smfrac{2\pi y}{M})}
\right|^2  C(\phi -\smfrac{2\pi y}{M})  }}}\\
& = & \frac{1}{2\pi M} \sum_{y=0}^{M-1}
{{\int_{\phi '=0}^{2\pi} d\phi ' {
\left|\sum_{j=0}^N\alpha_j \E^{\I j\phi '}
\right|^2  C(\phi ')  }}}\\
& = & \frac{1}{2\pi}{{\int_{\phi '=0}^{2\pi} d\phi ' {
\left|\sum_{j=0}^N\alpha_j \E^{\I j\phi '} \right|^2
C(\phi ')  }}},
\end{eqnarray*}
where $\phi '=\phi-\smfrac{2\pi y}{M}$. The expected cost does
not depend on $M$ as long as $M\ge N+1$.

Recall that we mentioned in the introduction that as $M \rightarrow
\infty$ the difference between the optimal $\bar{C}$ assuming the
inference rule $y \mapsto \tilde{\phi}_y = \frac{2\pi y}{M}$ and the
optimal $\bar{C}$ without such an assumption is $\epsilon_M
\rightarrow 0$. Since we have just shown that for all $M \geq N+1$,
the expected cost $\bar{C}$ is constant, this means that the
difference $\epsilon_M$ is in fact $0$ once $M \geq N+1$.  In other
words, assuming $\tilde{\phi}_y = \frac{2\pi y}{M}$ does not cost us
anything as long as we use $M \geq N+1$.

It is clear that the exact value of the expected cost now depends
only on the $\alpha_j$ values, that is, on the initial state, which
means that given a specific cost function all we have to do is to
choose an optimal initial state.

We emphasise that the schemes in Figs. \ref{optimal_fig} and
\ref{general_phase_est_fig} provide an optimal covariant estimation
scheme even for general cost functions not necessarily of the Holevo
class. Indeed, in \cite{DDEMMb} it is proved that there exists a
discrete POVM achieving the same average cost of any covariant
continuous POVM. By the Naimark theorem \cite{helstrom} this means
that there exists an orthogonal measurement on the system and an
ancilla achieving the discrete POVM, and a further extension allows
to have the POVM as rank-one (projectors of rank $r>1$ can be easily
connected to rank-one projectors via a suitable controlled-unitary
interaction with an ancilla).

Let us now address the problem of optimal input states for two
different cost functions. First we look at the minimization of the
``$1-$Fidelity'' cost function
$C_F(\phi,\tilde{\phi}_y)=\sin^2[(\phi-\tilde{\phi}_y)/2]$. The
minimum cost is achieved with
 the initial state
\begin{eqnarray}
\ket{x_N^{\mathrm{optimal}}} & = &
\sum_{j=0}^{N}{\sqrt{\smfrac{2}{N+2}}\sin\left({
\smfrac{(j+1)\pi}{N+2} }\right)\ket{j}}.
\end{eqnarray}
The error in fidelity of this protocol goes to zero according to
the square of the number of black boxes used $\bar{C}_F =
O(1/N^2)$. It is interesting to note that the fidelity of the more
conventional approach to phase-rotation estimation with the
uniform initial state ($\alpha_j=\smfrac{1}{\sqrt{N+1}}$ for all
$j$) only tends to zero \emph{linearly\/} in $N$. That is
$\bar{C}_F = \Omega(1/N)$.

Another cost function that is commonly used is the window function
that allows any error smaller than $\delta$:
$C_W^\delta(\phi,\tilde{\phi})=0$ if $|\phi-\tilde{\phi}|<\delta$,
but $C_W^\delta(\phi,\tilde{\phi})=1$ if
$|\phi-\tilde{\phi}|\geq\delta$. The minimisation of this cost leads
to optimal states with amplitudes $\alpha_j = 1/\sqrt{N+1}$, which
corresponds to what is effectively used by Shor's algorithm
\cite{shor, kitaev, CEMM}, and provides an expected cost in
$O(\frac{1}{\delta N})$.

In this paper we have addressed the general problem of finding the
optimal estimating procedure for the real parameter $\phi$ given $N$
copies of the single qubit phase rotation $u_{\phi}$ within a
general quantum circuit in finite dimensions. We considered the
general case where the circuit consists of an arbitrary input state
followed by any arrangement of the $N$ phase rotations interspersed
with arbitrary quantum operations. The main result was the proof
that in all cases, and for any covariant cost function, we want to
use the optimal phase estimation procedure is equivalent to a
quantum Fourier transform in an appropriate basis.

Our result is very general, and gives a recipe for finding the best
achievable phase estimation for a given cost function. In practice,
once we know the minimum cost possible, one can also search for and
use easier-to-implement phase estimation procedures that achieve the
same, or similar expected cost. Due to the generality of our main
result, it will surely find many other interesting applications in
physical and computational scenarios.

This is the first application of the polynomial method to
``black-boxes'' encoding continuous variables, in this case, one
real parameter. The method can also be applied to several real
parameters, as well as combinations of real and discrete parameters.


\begin{references}
\bibitem{nota}
This class of mixed states $\theta$ is characterised by the
restriction that the elements of $\theta$ have fixed phase along
the diagonals, i.e. $\theta_{hk}=|\theta_{hk}|e^{i\phi_{|h-k|}}$.
In this case the optimal $\chi$ is given by
$\chi_{hk}=e^{-i\phi_{|h-k|}}$.
\end{references}
\end{document}